\documentclass[5p,preprint]{elsarticle}
\biboptions{sort&compress}
\usepackage{lineno,isotope,mathrsfs}
\modulolinenumbers[5]
\usepackage{amsmath,amssymb,mathptmx,epsfig,bm,mathrsfs,feynmp}
\usepackage{color,slashed,nicefrac,exscale,multirow,soul,appendix}
\usepackage{times,txfonts,xcolor,graphicx,gensymb,tikz,tablefootnote}
\usepackage[normalem]{ulem}
\usepackage[breaklinks,colorlinks,citecolor=blue]{hyperref}
\definecolor{red}{rgb}{0.8,0,0}
\definecolor{violet}{rgb}{0.4,0,0.4}
\definecolor{green}{rgb}{0,0.5,0.0}
\definecolor{navy}{rgb}{0.0,0.0,0.6}
\definecolor{orange}{rgb}{0.8,0.2,0.0}

\newcommand{\bea}{\begin{eqnarray}}
\newcommand{\eea}{\end{eqnarray}}

\begin{document}
\begin{frontmatter}
\title{Bayesian constraints on covariant density functional 
equations of state of compact stars with new NICER mass-radius measurements}
\author[a]{Jia-Jie Li}
\ead{jiajieli@swu.edu.cn}
\author[a]{Yu Tian}
\author[b,c]{Armen Sedrakian}
\ead{sedrakian@fias.uni-frankfurt.de}
\address[a]{School of Physical Science and Technology, 
Southwest University, Chongqing 400715, China}
\address[b]{Frankfurt Institute for Advanced Studies, 
D-60438 Frankfurt am Main, Germany}
\address[c]{Institute of Theoretical Physics, University of Wroclaw,
50-204 Wroclaw, Poland}
\begin{abstract}
Recent advancements in astrophysical observations of compact stars, particularly
the new and updated NICER constraints, have provided mass-radius ($M$-$R$)
data for pulsars spanning masses from 1 to $2\,M_{\odot }$. These data offer
a unique opportunity to test modern theories of dense matter using multi-messenger
constraints. Covariant density functional (CDF) models of nuclear matter,
which capture a broad range of nuclear and astrophysical phenomena, provide
a robust theoretical framework to interpret these observations. This study
applies the Bayesian framework to a class of CDF models with density-dependent
meson-nucleon couplings, specifically those based on nucleonic degrees
of freedom. By incorporating the latest multi-messenger constraints, we
impose tighter limits on the parameter space of these models and assess
their consistency with observational data. Our analysis advances previous
efforts by refining the density-dependence parameterization and integrating
recent $M$-$R$ ellipses. This enables more stringent evaluations of dense
matter models in light of new astrophysical observations.
\end{abstract}
\begin{keyword}
Equation of state \sep Compact stars \sep 
Covariant density functional \sep Bayesian inference  
\end{keyword}
\end{frontmatter}
%
\section{Introduction}
\label{sec:Intro}
Recently, new NICER astrophysical constraints have been released for two
pulsars --- one canonical-mass $1.4\,M_{\odot}$ star J0437-4715 (hereafter
J0437,~\cite{Choudhury:2024}), and the one-solar-mass star PSR J1231-1411
(J1231, \cite{Salmi:2024b}). Combining these with two previously reported
data --- the two-solar-mass pulsar PSR J0740+6620 (J0740,~\cite{Riley:2021,Miller:2021})
and the canonical-mass star PSR J0030+0451 (hereafter J0030,~\cite{Riley:2019,Miller:2019}),
the mass range covered by the current data spans the range 1 to
$2\,M_{\odot}$. This data opens an unprecedented opportunity to explore
models of modern theories of dense matter subject to astrophysical constraints.
Notably, recent developments include the analysis of PSR J1231 which resulted
in two different ellipses in the mass-radius ($M$-$R$) plane, depending
on the radius prior~\citep{Salmi:2024b} and the reanalysis of PSR J0030,
which resulted in three different ellipses in the $M$-$R$ plane, each
corresponding to a different analysis method~\citep{Vinciguerra:2024}.

The covariant density functional (CDF) models of nuclear matter provide
a rigorous framework to address the full range of available data on nuclear
systems, ranging from the atomic chart to the astrophysics of compact stars
(CSs), for reviews 
see~\cite{Vretenar:2005,Niksic:2011,Oertel:2017,Piekarewicz:2019,Sedrakian:2023}.
These models were instrumental in addressing successfully such astrophysics
problems as hyperon puzzle in conjunction with two-solar mass CSs and tidal
deformability (TD) inference of GW170817 event. The CDF models are separated
into broad classes which (a) have non-linear meson contributions to the
effective Lagrangian and (b) keep only linear coupling but impose density-dependence
of the coupling, which captures the medium modifications of the meson-nucleon
vertices, see reviews~\cite{Oertel:2017,Sedrakian:2023} for discussions.

The Bayesian framework of constraining the properties of dense
matter in CSs given the observational constraints --- typically ellipses
in the $M$-$R$ diagram --- has attracted recently significant attention.
It allowed one to find correlations between the nuclear observables consistent
with the astrophysical inferences. The Bayesian framework has be applied
to models covering the range from fully physics agnostic non-parametric
models~\citep{Raaijmakers:2019,Landry:2020,Raaijmakers:2020,Legred:2021,
Pang:2021,Altiparmak:2022,Annala:2022,Annala:2023,Chimanski:2022,Rutherford:2024}
to microscopic models based on nuclear potentials~\citep{Mondal:2023,
Zhou:2023,Beznogov:2024a,Beznogov:2024b,Tsang:2024}
to density functional method-based CDF~\citep{Traversi:2020,Malik:2022,
Sun:2022,Zhu:2023,Beznogov:2023,Malik:2023,Salinas:2023,Huang:2024,Parmar:2024}.

The aim of this paper is to apply the Bayesian framework to CDF models
that have been developed 
in~\cite{Typel:1999,Lalazissis:2005,Lijj:2019a,Lijj:2019b,Lijj:2023b},
all of which correspond to the class of models of CDF with linear meson-baryon
couplings and density-dependent coupling constants. The utility of the
CDF framework, when compared to agnostic models, lies in its ability to
directly access the composition and quasiparticle spectra of constituents.
These are essential for physical applications, such as studies of transport
and neutrino interactions. Compared to microscopic models, CDFs enable
efficient exploration of parameter spaces at reduced numerical cost. We
aim to assess the compatibility of the parameter space over which these
models are defined with the recent multi-messenger observations of CSs.
Previous studies of this sort used simplified functions for density
dependence of meson-baryon couplings~\citep{Malik:2022,Beznogov:2023,Parmar:2024}.
Our analysis is aimed at revealing more stringent constraints on the CDF
models given that the most recent astrophysical inferences have not been
included in the Bayesian framework. Specifically, we will suggest and scrutinize
different scenarios that incorporate different combinations of the above-listed
$M$-$R$ ellipses. Previous studies employed simplified exponential density
dependencies for the $\sigma $- and $\omega $-meson fields, which are not
aligned with the standard formulations typically used in these models.
To address this inconsistency, the current work retains the established
density-dependent CDF forms as previously applied in the literature.

The present study is limited to matter with nucleonic degrees of freedom, 
which eliminates the possibility of nucleation of heavy baryons (hyperons 
and $\Delta$-resonances)~\citep{Lijj:2019b,Sedrakian:2023}, as well as quark
deconfinement~\citep{Alford:2008,Sedrakian:2023Parti}, at typically several
times the nuclear saturation density. Such a restriction allows us to focus
our analysis on a smaller set of parameters, which otherwise would contain
the hyperonic couplings within our CDF approach, or parameters of quark
matter equation of state (EOS) in any particular model of QCD. The impact
of non-nucleonic degrees of freedom on CS observables has been extensively
studied. For example, hyperonization reduces the maximum mass of CSs, therefore
the correlations sensitive to the high density physics, such as those involving
skewness of EOS will be affected~\citep{Lijj:2018,Lijj:2023a}. Furthermore, a
non-monotonic speed of sound is obtained within such models. Although definitive 
proof is not yet available, several recent studies based on physics-agnostic 
treatments of high-density matter~\citep{Altiparmak:2022,Gorda:2023,Yaonx:2024}
suggest a decrease in the speed of sound once the non-nucleonic degrees
of freedom nucleate. By construction, such behavior is not accounted
for within the present CDF framework

\section{CDF for nucleonic matter}
\label{sec:Model}
We use the CDF approach based on the Lagrangian of
stellar matter with nucleonic degrees of freedom
$ \mathscr{L} = \mathscr{L}_N + \mathscr{L}_m + 
\mathscr{L}_l,$
where the nucleonic Lagrangian is given by
\begin{align}
\label{eq:Lagrangian}
\begin{split}
\mathscr{L}_N  = 
\sum_N\bar\psi_N \Big[\gamma^\mu
\big(i \partial_\mu - g_{\omega}\omega_\mu 
    - g_{\rho} \bm{\tau} \cdot \bm{\rho}_\mu \big) \\
    -\big(m_N - g_{\sigma}\sigma\big) \Big]\psi_N,
\end{split} 
\end{align}
where $\psi _{N}$ are the nucleonic Dirac fields with mass $m_{N}$,
$\sigma,\,\omega _{\mu}$, and $\bm{\rho}_{\mu}$ are the mesonic fields
that mediate the interactions among the nucleon fields. The remaining pieces
of the Lagrangian correspond to the mesonic and leptonic contributions,
respectively.

The meson-nucleon couplings, which are density-dependent and are given
by
\begin{align}
g_{m}(\rho)=g_{m}\,(\rho_{\rm {sat}})f_m(r),
\end{align}
where index $m$ refers to mesons, the coupling constant
$g_{m}(\rho _{\mathrm{{sat}}})$ is given at saturation density
$\rho _{\mathrm{{sat}}}$, and the function $f_{m}(r)$ depends 
on the ratio $r=\rho /\rho _{\mathrm{{sat}}}$. For the isoscalar 
mesons
\begin{align}\label{eq:isoscalar_coupling}
f_{m}(r)=a_m\frac{1+b_m(r+d_m)^2}{1+c_m(r+d_m)^2}, \quad 
m = \sigma,\omega,
\end{align}
with conditions
$f_{m}(1)=1, f^{\prime\prime}_{m}(0)=0$, and 
$f^{\prime\prime}_{\sigma}(1)=f^{\prime\prime}_{\omega}(1)$,
which reduce the number of free parameters.
The density dependence for the isovector meson is 
taken in an exponential form:
\begin{align}\label{eq:isovector_coupling}
f_{\rho}(r) = e^{-a_\rho (r-1)}.
\end{align}

If we fix in the Lagrangian~\eqref{eq:Lagrangian} the nucleon and meson
masses to be (or close to) the ones in the vacuum then properties of infinite
nuclear matter can be computed uniquely in terms of seven adjustable parameters.
These are the three coupling constants at saturation density ($g_{
\sigma},\,g_{\omega},\,g_{\rho}$), and four parameters ($a_{\sigma},
\,d_{\sigma},\,d_{\omega},\,a_{\rho}$) that control their density dependences.
We consider uniform prior distributions of these seven CDF parameters within
reasonable intervals.

\section{Inference framework}
\label{sec:Bayesian}
The Bayesian analysis has been used in a variety of different research
fields to infer the probability distribution of unknown parameters in 
a model by exploiting the information from observation. This is 
accomplished by relying on the Bayes' theorem,
\begin{align}
p\,(\bm\theta\vert \bm d) = 
\frac{\mathcal{L}(\bm d \vert \bm\theta)\,p(\bm\theta)}
{\int\mathcal{L}(\bm d\vert\bm\theta)\,p(\bm\theta)\,d\bm\theta},
\end{align}
where available knowledge on model parameters $\bm \theta $ is expressed
as a prior distribution $p(\bm \theta )$, by combining with observational
data $\bm d$ in terms of likelihood functions
$\mathcal{L}(\bm d \vert\bm \theta)$. The posterior distributions
$p(\bm \theta \vert \bm d)$ are then updated with the information from
observables. The denominator is a normalization factor and acts as the
evidence of data. Below, we present a brief discussion on the constraints
and corresponding likelihoods we adopted in the present analysis.

\begin{table}[tb]
\centering
\caption{Symmetric nuclear matter (SNM) characteristics at saturation 
density~\citep{Lijj:2019a} and pure neutron matter (PNM) properties 
from $\chi$EFT computation~\citep{Hebeler:2013} that constrain the CDF 
parameters. The prior distributions assumed for the quantities are either 
a Gaussian distribution (G) or a Uniform distribution (U).
} 
\setlength{\tabcolsep}{7.8pt}
\label{tab:Nuclear_matter}
\begin{tabular}{cccccc}
\hline\hline
    &  Quantity        & Unit       & Interval          & Prior \\
\hline                  
    & $\rho_{\rm sat}$ & fm$^{-3}$  & $  0.153\pm 0.005$&   G   \\
    & $M^\ast_{\rm D}$ & $m_{\rm N}$& $  0.60 \pm 0.05$ &   G   \\
    & $E_{\rm sat}$    & MeV        & $-16.1  \pm 0.2$  &   G   \\
SNM & $K_{\rm sat}$    & MeV        & $230    \pm40 $   &   G   \\
    & $Q_{\rm sat}$    & MeV        & $[-1000, 1500]$   &   U   \\
    & $J_{\rm sym}$    & MeV        & $ 32.5  \pm 2.0$  &   G   \\
    & $L_{\rm sym}$    & MeV        & $[0, 100]$        &   U   \\
\hline                  
\multirow{2}*{PNM}&$P(\rho)$       & MeV/fm$^{3}$& N$^3$LO & G  \\
                  &$\epsilon(\rho)$& MeV/fm$^{3}$& N$^3$LO & G  \\
\hline\hline
\end{tabular}
\end{table}

\begin{figure*}[tb]
\centering
\includegraphics[width = 0.85\textwidth]{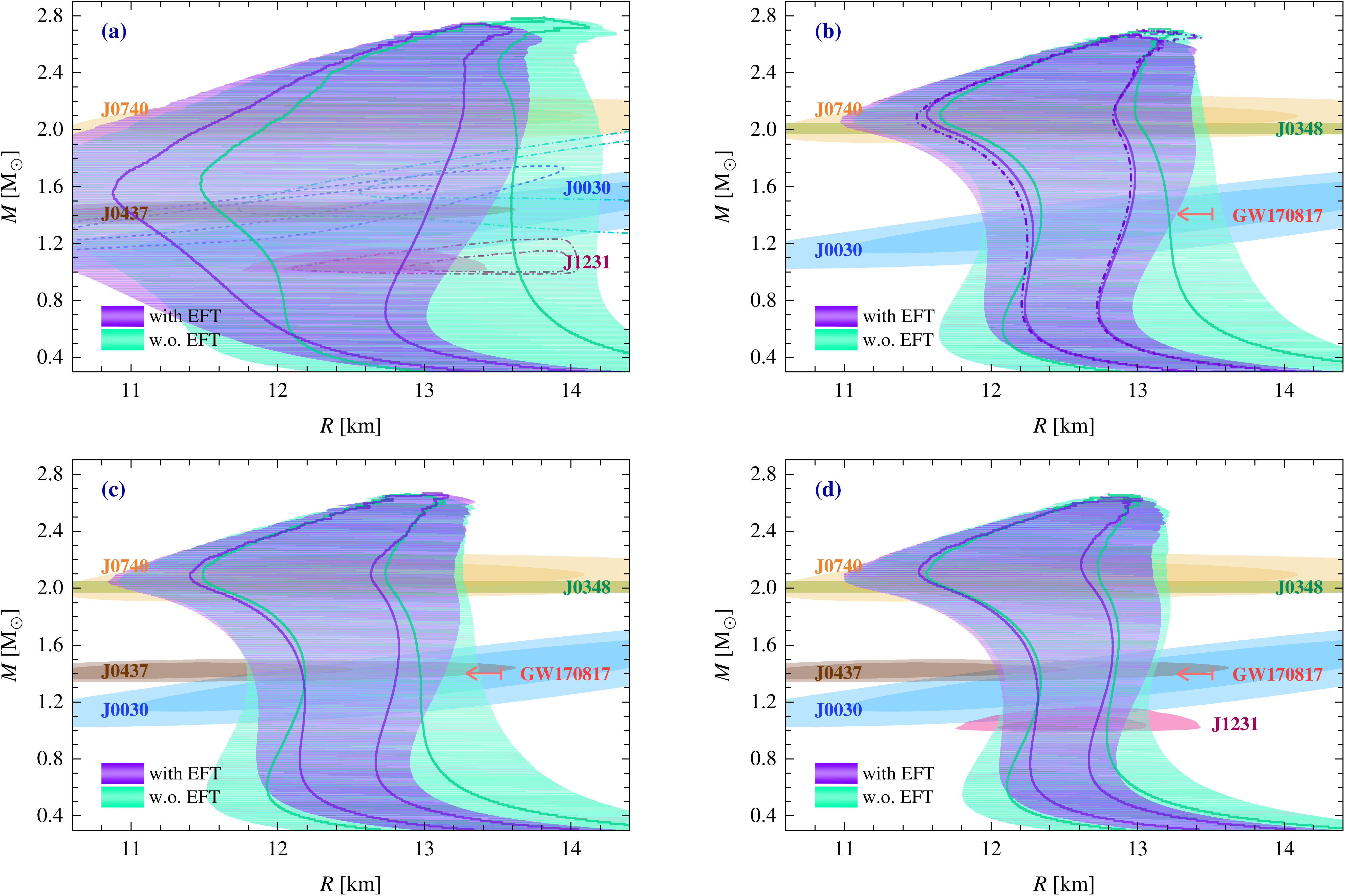}
\caption{ Posteriors for the mass-radius distribution under various
  constraints explored in this study are presented. The shaded regions
  represent the 95.4\% credible intervals (CI), while the lines
  indicate the 68.3\% CI.  Panel (a) imposes constraints solely on
  low-density nucleonic matter, while panels (b), (c), and (d)
  incorporate additional astrophysical constraints. These include mass
  measurements from the massive pulsar J0348, tidal deformability
  constraints derived from the GW170817 and GW190425 events, and
  mass-radius estimates obtained from NICER observations. Specifically, 
  panel (b) utilizes data from J0740 and J0030, panel (c) adds data 
  from J0437 to this ensemble, and panel (d) further incorporates data 
  from J1231. For comparison, the impact of the $\chi$EFT constraint 
  is highlighted in each panel. Panel (a) also provides an overview of 
  new and updated NICER results, including the latest estimate for 
  J0437 from~\cite{Choudhury:2024}, two estimates for J1231
  from~\cite{Salmi:2024b}, the reanalysis of J0740
  from~\cite{Salmi:2024a}, and three possible solutions for J0030
  from~\cite{Vinciguerra:2024}. Panel (b) includes results computed
  using previous NICER data from~\cite{Riley:2021} and~\cite{Riley:2019} 
  for comparison.}
\label{fig:MR_EXample}
\end{figure*}

\begin{table*}[tb]
\caption{
New astrophysical constraints are used for the scenarios in the 
present work.}
\setlength{\tabcolsep}{9.4pt}
\label{tab:Scenarios}
\begin{tabular}{ccccccccccccccc}
\hline\hline
\multirow{2}*{Scenario}   & \multirow{2}*{J0348} & GW $^{\rm a}$ & J0740 & J0437 &
\multicolumn{3}{c}{J0030 $^{\rm b}$} & \multicolumn{2}{c}{J1231 $^{\rm c}$} \\
\cline{6-8} \cline{9-10}
                       &      &(2)& ST-U &CST+PDT& ST+PST & ST+PDT & PDT-U  &  PDT-U (i)  &  PDT-U (ii) \\
\hline
Baseline&$\times$ & $\times$& $\times$&         & $\times$&         &         &         &          \\
\hline
A       &$\times$ & $\times$& $\times$& $\times$& $\times$&         &         & $\times$&          \\
B       &$\times$ & $\times$& $\times$& $\times$&         & $\times$&         & $\times$&          \\
C       &$\times$ & $\times$& $\times$& $\times$&         &         & $\times$& $\times$&          \\
\hline
D       &$\times$ & $\times$& $\times$& $\times$& $\times$&         &         &         & $\times$ \\
E       &$\times$ & $\times$& $\times$& $\times$&         & $\times$&         &         & $\times$ \\
F       &$\times$ & $\times$& $\times$& $\times$&         &         & $\times$&         & $\times$ \\
\hline\hline
\end{tabular}
\footnotesize{${\rm ^a}$ Both the confirmed two (most likely) binary 
neutron star mergers GW170817~\citep{LIGOScientific:2017} and 
GW190425~\citep{LIGOScientific:2020a} are considered.} \\
\footnotesize{${\rm ^b}$ For PSR J0030, the ST+PST refers to the NICER-only 
analysis of the same data set from~\cite{Riley:2019} with an improved analysis 
pipeline and setting, ST+PDT and PDT-U are two modes preferred in the joint 
analysis of NICER and XMM data for which the ST+PDT results are more consistent 
with the magnetic field geometry inferred for the gamma-ray emission for this 
source, and the PDT-U is the most complex model and is preferred by the Bayesian 
evidence~\cite{Vinciguerra:2024}.} \\
\footnotesize{${\rm ^c}$ For PSR J1231, the PDT-U (i) refers to the model 
that used an informative radius prior based on the results 
of~\cite{Raaijmakers:2021}, and (ii) the one that limited the 
radius between 10 and 14~km.} 
\end{table*}

\begin{figure*}[tb]
\centering
\includegraphics[width = 0.85\textwidth]{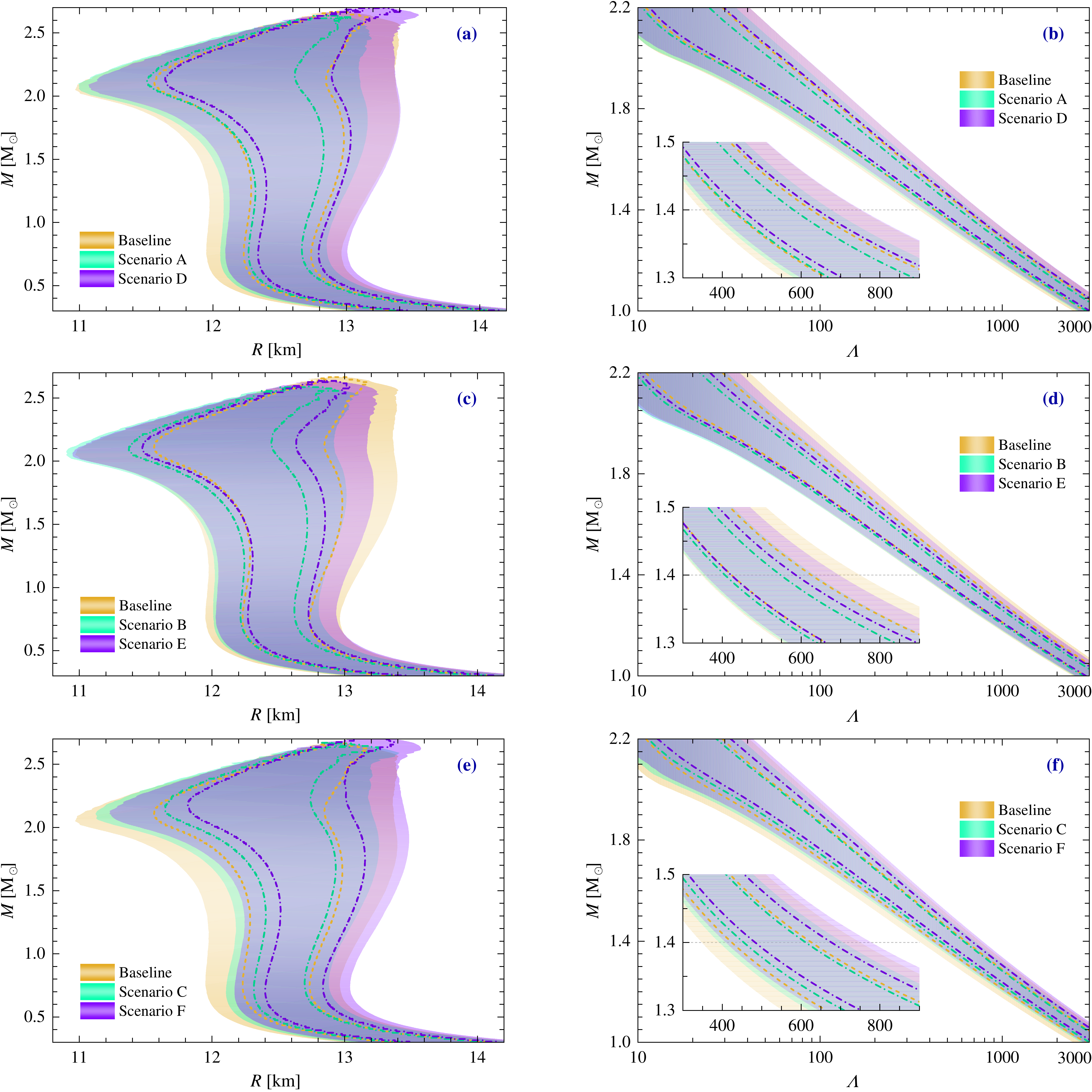}
\caption{
Posteriors for mass-radius (left panels) and mass-tidal 
deformability (right panels) distributions under seven scenarios
explored in this work. The shaded regions represent the distributions 
at 95.4\% CI, while the lines for results at 68.3\% CI. 
In the right panels, the insets magnify the TDs for typical compact stars.
}
\label{fig:MR_MLambda}
\end{figure*}

\subsection{Low-density nuclear matter properties}
The energy density of isospin asymmetric matter is customarily 
split into an isoscalar and an isovector term:
\begin{align}
\label{eq:isospin_expansion}
E\,(\rho, \delta) \simeq 
E_0\,(\rho) + E_{\rm sym}\,(\rho)\,\delta^2 + {\mathcal O}\,(\delta^4)
\end{align}
where $\rho= \rho_n + \rho_p$ is the baryonic density, with 
$\rho_{n(p)}$ denoting the neutron (proton) density, 
$\delta = (\rho_{n}-\rho_{p})/\rho$ is the isospin asymmetry, 
and $E_0(\rho)$ and $E_{\rm sym}(\rho)$ are, respectively, 
the energy of symmetric matter and the symmetry energy. 
At densities close to the saturation $\rho_{\rm sat}$ 
Eq.~\eqref{eq:isospin_expansion} can be further Taylor 
expanded as
\begin{equation}
\label{eq:Taylor_expansion}
\begin{split}
E\,(\rho, \delta) 
& \simeq 
E_{\rm{sat}} + \frac{1}{2!}K_{\rm{sat}}\chi^2
+ \frac{1}{3!}Q_{\rm{sat}}\chi^3 \\
& + \,J_{\rm{sym}}\delta^2 + L_{\rm{sym}}\delta^2\chi 
+ \frac{1}{2!}K_{\rm{sym}}\delta^2\chi^2
+ {\mathcal O}\,(\chi^4, \chi^3\delta^2),
\end{split}
\end{equation}
where $\chi =(\rho -\rho _{\mathrm{{sat}}})/3\rho _{\mathrm{{sat}}}$. 
The coefficients of the expansion are known as 
\textit{nuclear matter characteristics at saturation
density}, namely, \textit{incompressibility} $K_{\mathrm{{sat}}}$, the 
\textit{skewness} $Q_{\mathrm{{sat}}}$, the \textit{symmetry energy} 
$J_{\mathrm{{sym}}}$ and its \textit{slope parameter} $L_{\mathrm{{sym}}}$. 
The coefficients of the leading-order terms in the expansion, namely 
$K_{\mathrm{{sat}}}$ and $J_{\mathrm{{sym}}}$ are well determined,
whereas the coefficients of higher-order terms, namely
$Q_{\mathrm{{sat}}}$ and $L_{\mathrm{{sym}}}$ are not well known. 
In addition to the parameters in Eq.~\eqref{eq:Taylor_expansion}, 
the Dirac mass $M^{\ast}_{\mathrm{D}}$ at saturation is well-constrained. 
This parameter plays a crucial role in the quantitative description of 
finite nuclei phenomena, such as spin-orbit splitting. For quantities 
known with an uncertainty of approximately 10\%, we adopt Gaussian priors, 
while for others, we use uniform priors. The mean values, standard deviations, 
or intervals for each parameter are detailed in Table~\ref{tab:Nuclear_matter}. 
Notably, broader ranges are considered for $Q_{\mathrm{{sat}}}$ and 
$L_{\mathrm{{sym}}}$ to encompass the variations reported in different 
studies.

In addition to the constraints above, we incorporate results for pure neutron
matter up to around nuclear saturation density based on chiral effective
field theory ($\chi $EFT) interactions. These constrain the low-density
regime of nucleonic EOS. In the present multi-messenger analyses, we use
the N$^{3}$LO constraints from~\cite{Hebeler:2013} and assume the uncertainty
band as a Gaussian distribution. As an illustration of the impact of this
constraint, we show in Fig.~\ref{fig:MR_EXample}~(a) the $M$-$R$ 
posterior distributions at 68.3\% and 95.4\% CIs for models considering only 
low-density nucleonic matter constraints.

\subsection{Astrophysical observations}
We now briefly describe our implementations of the likelihoods for 
various astrophysical observations. The product of the individual 
likelihoods of these sources finally gives the total likelihood.

\begin{itemize}
\item 
The \textit{NICER} collaborations have delivered the joint measurement
of mass and radius through pulse profile modeling of four millisecond pulsar:
a massive $\sim 2\,M_{\odot}$ star PSR J0740, two canonical-mass
$\sim 1.4\,M_{\odot}$ stars PSR J0030 and J0437, and a low mass
$\sim 1\,M_{\odot}$ star PSR J1231. We construct the likelihood function
for each of the sources using the Gaussian kernel density estimation (KDE)
with the released posterior $(M, R)$ samples $\bm S$,
\begin{eqnarray}
\mathcal{L}_{\rm NICER}\,(\bm\theta_{\rm{EOS}}) = {\rm KDE}\,(M, R|\bm{S}),
\end{eqnarray}
where the mass $M$ and radius $R$ for the star are functions of its central
pressure and the sampled EOS parameters $\theta _{\mathrm{{EOS}}}$. The model
samples of each pulsar we implement for the present analysis are listed
in Table~\ref{tab:Scenarios}.
\item 
To date, the \textit{gravitational wave} (GW) events 
GW170817~\citep{LIGOScientific:2017}
and GW190425~\citep{LIGOScientific:2020a} are the only confirmed binary
neutron star mergers detected during the LVK collaboration's observing
runs. In the analysis of a single GW event, the likelihood function used
for Bayesian inference depends on a parameter vector,
$\bm{\theta}_{\mathrm{GW}}$, which includes two key components: parameters relevant
for constraining the EOS of dense matter, $\bm{\theta}_{\mathrm{EOS}}$, and
nuisance parameters, $\bm{\theta}_{\mathrm{nuis.}}$, necessary for modeling
GW-emitting binaries. However, the inclusion of numerous nuisance parameters
--- often numbering in the dozens --- substantially slows down the sampling
process. To address this, the likelihood is computed using high-precision
interpolation in TOAST~\citep{Hernandez:2020}, which marginalizes over
these nuisance parameters to streamline the analysis. It can be written
\begin{eqnarray}
\mathcal{L}_{\rm GW}\,(\bm\theta_{\rm{EOS}}) = 
F\,(\mathcal{M},\,q,\,\Lambda_1,\,\Lambda_2),
\end{eqnarray}
where $\mathcal{M} = (M_{1}M_{2})^{3/5}/(M_{1}+M_{2})^{1/5}$ is the chirp
mass, $q = M_{1}/M_{2}$ is the mass ratio, and $\Lambda _{1}(M_{1})$ and
$\Lambda _{2}(M_{2})$ the TDs of the individual star.
\item
To approximate the mass measurements of \textit{massive pulsars} (MP),
e.g., PSR J0348+0432 (hereafter J0348,~\citep{Antoniadis:2013}), we use
Gaussian distributions, and the cumulative density function of Gaussian
is applied to build the likelihood,
\begin{eqnarray}
\mathcal{L}_{\rm MP}\,(\bm\theta_{\rm{EOS}}) = \frac{1}{2}
\left[1 + {\rm erf} \left(\frac{M_{\rm{max}}(\bm\theta_{\rm{EOS}}) - M}
{\sqrt{2}\sigma}\right)\right],
\end{eqnarray}
where ${\mathrm{erf}}\,(x)$ is the error function, $M$ and $\sigma $ are the
mean and the standard deviation of the mass measurements for the source,
respectively. We do not take into account the mass measurement for PSR
J0740~\citep{NANOGrav:2019,Fonseca:2021}, to avoid double counting with
its NICER estimate.
\end{itemize}

\section{Results and implications}
\label{sec:Results}

\begin{figure*}[tb]
\centering
\includegraphics[width = 0.85\textwidth]{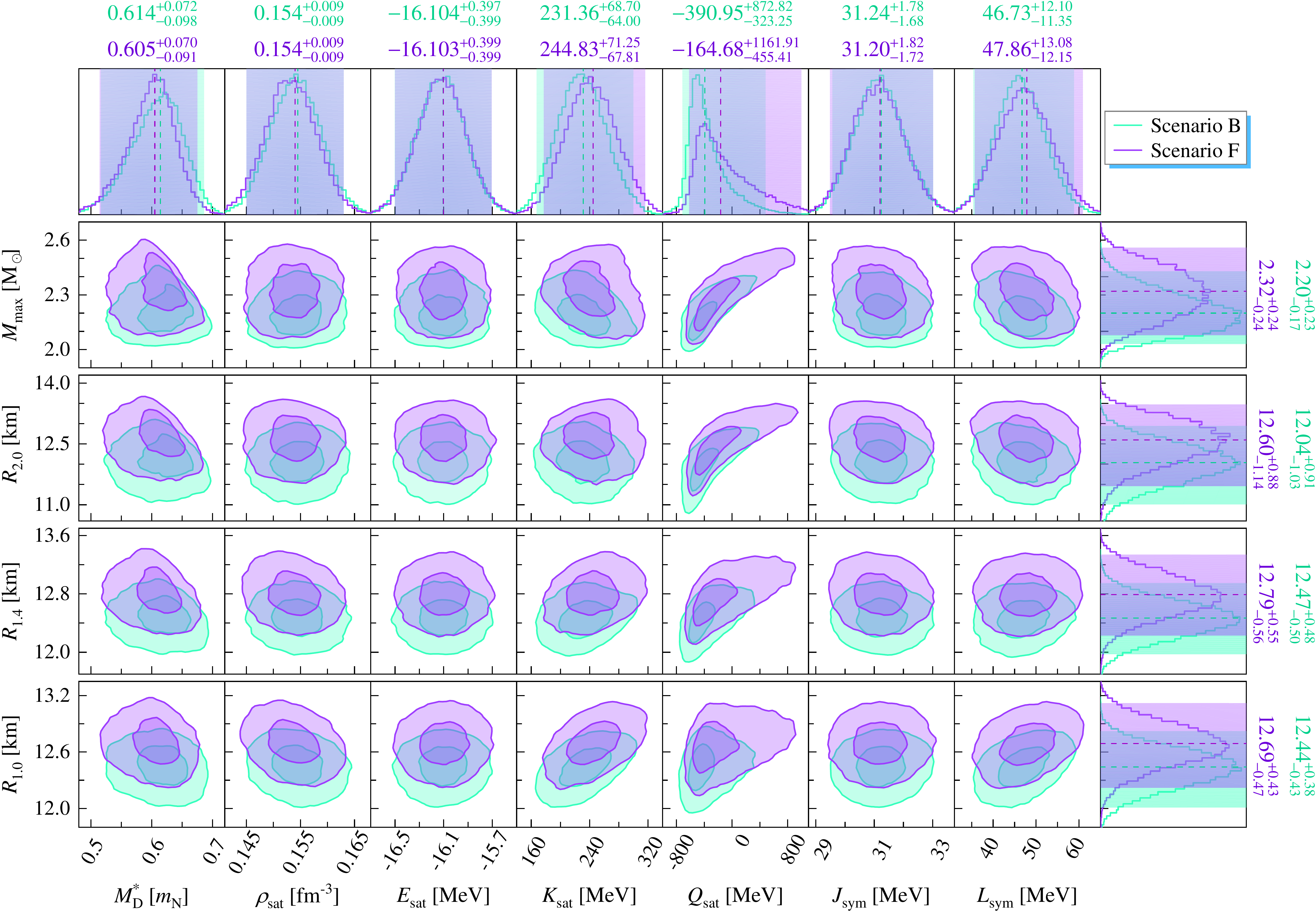}
\caption{
Posteriors for the correlations of characteristic parameters 
of symmetric nuclear matter at saturation and gross properties 
of compact stars under scenarios B and F which allow for 
the softest and stiffest models respectively. The light and dark 
shaded regions indicate respectively the 68.3\% and 95.4\% CI 
regions of the two-dimensional distributions. The one-dimensional 
posteriors for each quantity are given along the plots where 
vertical lines indicate the median positions and shaded intervals 
for -- 95.4\% CI, respectively. 
}
\label{fig:NN_Correlation}
\end{figure*}

\begin{figure*}[tb]
\centering
\includegraphics[width = 0.85\textwidth]{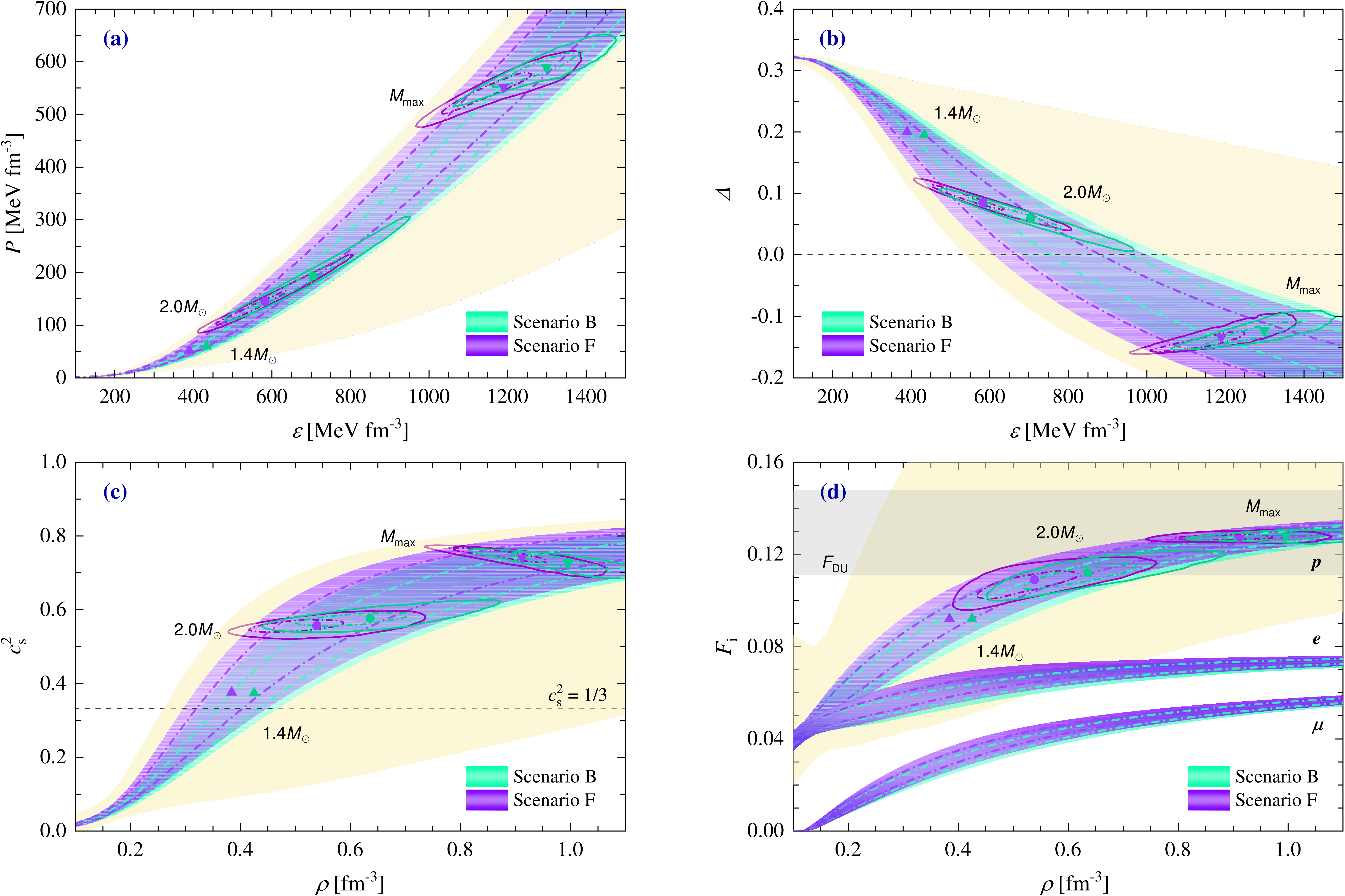}
\caption{ Posteriors for the nucleonic equation of state (panel a)
  distribution and dimensionless quantities characterizing the
  properties of dense matter -- trace anomaly $\Delta$ (panel b),
  sound speed squared $c_s^2$ (panel c), and the particle fraction
  (panel d) under scenarios B and F which allow for the softest and
  stiffest models respectively. The shaded regions represent the
  distributions at 95.4\% CI, while the lines for -- 68.3\% CI. In
  each panel, the contours show the corresponding posterior
  distributions and symbols of the median positions of the respective
  $1.4,\,2.0\,M_{\odot}$ stars and the maximum-mass configurations.
  In panel (d) the gray band labeled with $F_{\rm DU}$ frames the
  possible threshold values for the onset of direct Urca (DU) 
  cooling process. The 99.7\% CI regions after applying only the 
  SNM constraints at saturation density are incorporated as well 
  (in light yellow).
}
\label{fig:EOS_FRaction}
\end{figure*}

Our results are summarized in~Figs.~\ref{fig:MR_EXample}--\ref{fig:EOS_FRaction}.
We apply the astrophysical and theoretical constraints both selectively
and in concert to elucidate the importance of their individual impact as
well as the overall picture that emerges from a combination of such constraints.
We first discuss the roles of constraints individually.

\subsection{Impact of chiral EFT constraint}
Among the families of CDF parameterization of neutron matter of the EOS
of dense matter, only those are compatible with the $\chi $EFT which are
soft at low densities and remain such up to $2$-$3\,\rho _{\mathrm{sat}}$. More
quantitatively, the CDFs that are compatible with $\chi $EFT have the range
of symmetry energy and its slope at nuclear saturation density limited
by $29.5 \leq J_{\mathrm{sym}} \leq 33$~MeV and
$35 \leq L_{\mathrm{sym}} \leq 60$~MeV (at 95.4\% CI), respectively~\citep{Lijj:2019b}.
This leads to small uncertainty bands for radii of sub-canonical-mass stars,
as illustrated in Fig.~\ref{fig:MR_EXample}. As these constraints apply
only at low density, it is evident that the density dependence of isovector
quantities $J_{\mathrm{sym}}$ and $L_{\mathrm{sym}} $ have little impact on the properties
of high-mass CSs. This is clearly seen in panels (b-d) of Fig.~\ref{fig:MR_EXample}
where the posterior distributions for computations with and without
$\chi $EFT constraint are almost identical for sequences with
$M > 1.4\,M_{\odot}$.

\subsection{Impact of individual astrophysical source}
The masses of massive pulsars PSR J0348 and J0740 place strict constraints
on the high-density behavior of nucleonic EOS -- characterized by the value
of $Q_{\mathrm{sat}}$ in a CDF model -- which is set to
$Q_{\mathrm{sat}} \gtrsim -700$~MeV. For lower values of this parameter, the
masses of these pulsars are not reproduced. Furthermore, they impact the
lower value of the radius of stars, which is clearly visible by comparing
 panels (a) and (b) in Fig.~\ref{fig:MR_EXample}. Including these objects
results in the lower limit of the radius of a $2\,M_{\odot}$ star
$\sim 11$~km at 95.4\% CI. Combining with the lower limit on the value of
$L_{\mathrm{sym}}$ (constrained by $\chi $EFT) one finds for the radius of a
canonical-mass ($1.4\,M_{\odot}$) star $R_{1.4} \gtrsim 12$~km. On the
other hand, the ellipses for PSR J0348 and J0740 do not limit the radii
from above significantly. More stringent constraint on this limit comes
from TDs derived from GW170817 event, which place an \textit{upper limit}
on the radii of canonical-mass stars $R_{1.4} \lesssim 13.5$~km. Furthermore,
we have checked the posteriors by replacing the updated NICER estimates
with previous data from~\cite{Riley:2021} and~\cite{Riley:2019}, the difference
for boundaries is well within 0.1~km. This further strengthens the importance
of maximum mass and GW constraints for dense matter modeling.

In Fig.~\ref{fig:MR_EXample}~(c), the $M$-$R$ ellipse derived for PSR J0437
significantly overlaps with the $M$-$R$ inferences from GW170817 
event~\citep{LIGOScientific:2020a},
which reinforces the selection of EOS based solely on GW data or PSR J0437
data. Indeed, as shown in Fig.~\ref{fig:MR_EXample}~(b, c), inclusion 
of PSR J0437 inference in addition to GW data shifts the $M$-$R$ distributions
towards a lower radius but only by $\sim 0.2$~km.

Once the $M$-$R$ estimate from the PDT-U (i) model for PSR J1231 is included,
the posterior region is narrowed noticeably by 0.2~km from both sides;
see Fig.~\ref{fig:MR_EXample}~(d). This narrowing is because the J1231
PDT-U (i) results are consistent with the inferred radius of J0437. However,
if the estimate from the PDT-U (ii) model which favors a larger radius
is included in the analysis, the upper limit for radius is relaxed by about
0.3~km; see Fig.~\ref{fig:MR_MLambda}~(a).

In closing, the mass measurements of PSR J0348 and J0740, along with the
TDs inferred from GW170817 and the $\chi $EFT calculations, provide fundamental
constraints for determining the basic posterior region of the $M$-$R$ distribution.
This posterior region shows significant overlap with the NICER estimates
for pulsars. Consequently, we designate the posterior shown in Fig.~\ref{fig:MR_EXample}~(b)
as the ``Baseline'' for comparison with the full posteriors presented in
Fig.~\ref{fig:MR_MLambda}, which incorporate NICER estimates for all four
pulsars.

\subsection{Implications from multimessenger data}
Fig.~\ref{fig:MR_MLambda} summarizes the posteriors for $M$-$R$ and
$M$-$\Lambda $ distributions under seven different scenarios A - F and
``Baseline'' defined in Table~\ref{tab:Scenarios}. Overall the 
posterior distributions for $M$-$R$ feature similar shapes. The posterior 
distributions narrow somewhat compared to the ``Baseline'' scenario, due 
to the combined effect of the PSR J0030 and J0437 NICER results, but most 
of the scenarios fully remain within the ``Baseline'' contours (with the 
exception of scenario F). The posterior distributions for TD naturally 
show similar trends as for the $M$-$R$ posteriors.

As anticipated, the new NICER estimates for PSR J0437 and J1231, along
with the reanalysis of PSR J0030, affect only the finer details of the
posterior compared to the ``Baseline'' scenario. The tightest credible
regions arise in scenario B, where more compact estimates for PSR J1231
(PDT-U (i) model) and J0030 (ST+PDT model), favoring softer EOS at densities
below 2 and $3\,\rho _{\mathrm{sat}}$, respectively, are used; see 
Fig.~\ref{fig:MR_MLambda} (b, c). In contrast, the widest credible 
regions are predicted in the scenario F, which incorporates less compact 
estimates for PSR J1231 (PDT-U (ii) model) and J0030 (PDT-U model); see 
Fig.~\ref{fig:MR_MLambda} (d, e).

Fig.~\ref{fig:NN_Correlation} illustrates the posterior distributions for
the correlations between the characteristic parameters of nuclear matter
at saturation density and the macroscopic properties of CSs under scenarios
B and F. The characteristic parameters generally follow Gaussian distributions,
except for the isoscalar skewness, $Q_{\mathrm{sat}}$. The correlations between
these parameters and the properties of CSs are similar across both scenarios.
The isoscalar skewness, $Q_{\mathrm{sat}}$, which influences both the maximum
masses and radii of CSs, plays a dominant role in determining the maximum
mass by effectively modifying the EOS at supra-saturation densities. Meanwhile,
the isovector slope, $L_{\mathrm{sym}}$, constrained by $\chi$EFT calculations
of low-density neutron matter, is narrowly distributed around
$L_{\mathrm{sym}} \sim 47$~MeV and has negligible impact on the overall properties
of massive CSs.

Scenario B favors softer EOS featuring lower values for the incompressibility,
$K_{\mathrm{sat}} = 231.4_{-33.3}^{+34.4}$~MeV and negative values for the skewness,
$Q_{\mathrm{sat}} = -391.0_{-170.2}^{+333.6}$~MeV (at 68.3\% CI); while scenario
F favors stiffer EOS featuring higher values for
$K_{\mathrm{sat}} = 244.8_{-34.7}^{+36.4}$~MeV and allows for positive value
for $Q_{\mathrm{sat}} = -164.7_{-280.1}^{+548.9}$~MeV (at 68.3\% CI). The value
of $Q_{\mathrm{sat}}$ can further reach up to 900~MeV at a 95.4\% CI. Note that
in our modeling, the parameters $K_{\mathrm{sat}}$ and $Q_{\mathrm{sat}}$ are independent
of each other, this is in contrast to previous studies of this type which
simplified the functions~\eqref{eq:isoscalar_coupling} for density dependence
of meson-baryon couplings~\citep{Malik:2022,Beznogov:2023,Parmar:2024}.

Thus, scenarios B and F represent two distinct combinations of the current
NICER estimates for pulsars, corresponding to the softest and stiffest
models, respectively. The radius for a canonical-mass star is
$R_{1.4} = 12.47_{-0.50}^{+0.48}$ km (at 95.4\% CI) for scenario B and
$12.79_{-0.56}^{+0.55}$ km for scenario F. The corresponding TD is
$\Lambda _{1.4} = 472_{-121}^{+163}$ for scenario B and
$571_{-162}^{+207}$ (at 95.4\% CI) for scenario F. The maximum mass is
$M_{\mathrm{max}} = 2.20_{-0.17}^{+0.23}\,M_{\odot}$ and
$2.32_{-0.24}^{+0.24}\,M_{\odot}$ respectively for scenarios B and F. The
latter scenario therefore allows for a static CS interpretation for the
secondary component of the GW190814 event~\citep{LIGOScientific:2020b}.

Fig.~\ref{fig:EOS_FRaction} shows the posterior for EOS distribution and
dimensionless quantities characterizing the properties of dense strongly
interacting matter -- the dimensionless trace anomaly
$\Delta = 1/3-P/\varepsilon $~\citep{Fujimoto:2022} (panel b), sound speed
squared $c_{s}^{2}$ (panel c) -- and the particle fraction (panel d) under
scenarios B and F, respectively. For each scenario, the median positions
of the respective 1.4, $2.0\,M_{\odot}$ and the maximum-mass configurations
are marked in the plots.

In Fig.~\ref{fig:EOS_FRaction}~(b), the trace anomaly $\Delta $ in the
innermost cores of massive CSs ($M \gtrsim 2.0\,M_{\odot}$) tends to approach
zero and takes negative values for maximum mass configurations. The lower
limits are predicted as $\Delta \geqslant -0.123^{+0.038}_{-0.027}$ (at 95.4\% CI)
for scenario B and $\Delta \geqslant -0.134^{+0.049}_{-0.023}$ for scenario
F. These results align with the constraints reported in~\cite{Musolino:2024},
which are derived using agnostic EOS models.

Fig.~\ref{fig:EOS_FRaction} (c) shows the ranges of sound speed squared
$c_{s}^{2}$ predicted by the EOS models, which is directly related to the
slope of the EOS. Observations of two-solar mass CSs suggest that the pressure
must rise rapidly with increasing energy density to counteract gravitational
collapse. On the other hand, it is well known that the so-called conformal
limit $c_{s}^{2}=1/3$ (in units of speed of light) must be reached at extremely
high densities, where interactions between ultra-relativistic quarks vanish
due to asymptotic freedom. Our results agree with the studies which indicate
that the bound $c_{s}^{2} \le 1/3$ is significantly violated
(see e.g., \cite{Yao:2023,Lijj:2023a,Fujimoto:2019,Tews:2018,Bedaque:2014}).
At the same time, it shows that the CDF models are capable of producing
massive CSs without violating the causality, i.e., $c_{s}^{2}>1$ which
is not the case for (some) non-relativistic approaches. Note also, that
the slope of the $c_{s}$ remains convex above saturation density, which
is in contrast to the case of hypernuclear matter treated in mean-field 
approximation, in which case the slope becomes concave at the point of 
onset of heavy baryons~\citep{Lijj:2018,Lijj:2023a}. It also contrasts 
the case where phase transition is allowed at densities relevant for CSs 
which could lead to a decrease in the speed of sound before reaching its 
asymptotically free value~\cite{Komoltsev:2024}.

In Fig.~\ref{fig:EOS_FRaction} (d) the gray band indicates the direct Urca
(DU) threshold, which is, however, model dependent~\citep{Klahn:2006}.
For $\mu^{-}$ free case the threshold value is 11.1\%; in the limit of
massless muons, which is applicable for high densities matter, it yields
an upper limit of 14.8\%. As seen in this plot, due to the softness of
the nucleonic EOS at low densities, the DU process will be mostly disallowed
in CSs with $M \leq 2\,M_{\odot}$ utilizing CDF EOS. However, note that
the DU threshold may be smoothed out due to short-range correlations in
dense matter~\citep{Sedrakian:2024}.

\begin{table*}[tb]
\caption{
Characteristic parameters of symmetric nuclear matter at saturation density 
from the posterior distributions for the different astrophysical scenarios. 
The upper and lower values correspond to the 68.3\% CI.}
\setlength{\tabcolsep}{7.0pt}
\label{tab:NM_Posterior}
\centering
\begin{tabular}{ccccccccc}
\hline \hline 
Par.             & Unit             & Baseline   & 
Scenario A       & Scenario B       & Scenario C & 
Scenario D       & Scenario E       & Scenario F \\
\hline 
$M_{\rm D}^\ast$&$m_{\rm N}$& ${0.609}_{-0.044}^{+0.038}$ & 
${0.611}_{-0.045}^{+0.038}$ & ${0.614}_{-0.047}^{+0.039}$ &
${0.608}_{-0.044}^{+0.037}$ & ${0.607}_{-0.045}^{+0.038}$ & 
${0.611}_{-0.046}^{+0.039}$ & ${0.605}_{-0.043}^{+0.035}$ \\
$\rho_{\rm sat}$& fm$^{-3}$ & ${0.154}_{-0.005}^{+0.005}$ &
${0.154}_{-0.005}^{+0.005}$ & ${0.154}_{-0.005}^{+0.005}$ & 
${0.154}_{-0.005}^{+0.005}$ & ${0.154}_{-0.005}^{+0.005}$ & 
${0.154}_{-0.005}^{+0.005}$ & ${0.154}_{-0.005}^{+0.005}$ \\
$E_{\rm sat}$   &  MeV      & ${-16.11}_{-0.20}^{+0.20}$  & 
${-16.11}_{-0.20}^{+0.20}$  & ${-16.10}_{-0.20}^{+0.20}$  & 
${-16.11}_{-0.20}^{+0.20}$  & ${-16.11}_{-0.20}^{+0.20}$  & 
${-16.11}_{-0.20}^{+0.20}$  & ${-16.10}_{-0.20}^{+0.20}$  \\
$K_{\rm sat}$   &  MeV      & ${233.9}_{-35.5}^{+37.1}$   & 
${234.1}_{-33.3}^{+35.4}$   & ${231.4}_{-33.3}^{+34.4}$   & 
${235.2}_{-33.3}^{+35.0}$   & ${242.8}_{-35.8}^{+36.1}$   & 
${239.1}_{-35.5}^{+36.3}$   & ${244.8}_{-34.7}^{+36.4}$   \\
$Q_{\rm sat}$   &  MeV      & ${-278.4}_{-237.7}^{+499.2}$& 
${-330.6}_{-200.2}^{+409.5}$& ${-391.0}_{-170.2}^{+333.6}$& 
${-249.8}_{-238.7}^{+469.7}$& ${-271.7}_{-228.7}^{+493.4}$& 
${-357.4}_{-182.8}^{+376.1}$& ${-164.7}_{-280.1}^{+548.9}$\\
$Z_{\rm sat}$   &  MeV      & ${4961}_{-2296}^{+1630}$    & 
${4888}_{-2340}^{+1574}$    & ${4931}_{-2330}^{+1595}$    & 
${4932}_{-2289}^{+1544}$    & ${4626}_{-2472}^{+1714}$    & 
${4652}_{-2529}^{+1718}$    & ${4641}_{-2508}^{+1663}$    \\
\hline
$J_{\rm sym}$   &  MeV      & ${31.18}_{-0.86}^{+0.88}$   & 
${31.23}_{-0.86}^{+0.88}$   & ${31.24}_{-0.86}^{+0.88}$   & 
${31.18}_{-0.85}^{+0.89}$   & ${31.22}_{-0.85}^{+0.90}$   & 
${31.27}_{-0.85}^{+0.88}$   & ${31.20}_{-0.88}^{+0.89}$   \\
$L_{\rm sym}$   &  MeV      & ${46.47}_{-5.98}^{+6.33}$   & 
${46.92}_{-5.82}^{+5.93}$   & ${46.73}_{-5.67}^{+5.94}$   & 
${46.73}_{-5.95}^{+6.04}$   & ${47.80}_{-6.05}^{+6.30}$   & 
${47.78}_{-5.97}^{+6.07}$   & ${47.86}_{-6.27}^{+6.46}$   \\
$K_{\rm sym}$   &  MeV      & ${-99.3}_{-17.3}^{+25.6}$   & 
${-102.6}_{-15.8}^{+23.2}$  & ${-106.0}_{-14.7}^{+21.0}$  & 
${-98.3}_{-16.9}^{+25.1}$   & ${-99.3}_{-16.9}^{+24.8}$   & 
${-104.5}_{-14.9}^{+21.6}$  & ${-94.2}_{-18.2}^{+26.6}$   \\
$Q_{\rm sym}$  &  MeV       & ${797.1}_{-173.5}^{+156.9}$ & 
${778.9}_{-163.1}^{+159.1}$ & ${776.1}_{-159.1}^{+155.5}$ & 
${792.1}_{-166.3}^{+158.2}$ & ${757.8}_{-175.1}^{+165.5}$ & 
${751.2}_{-166.1}^{+162.7}$ & ${765.3}_{-177.0}^{+166.5}$ \\
$Z_{\rm sym}$   &  MeV      & ${-5616}_{-2312}^{+1696}$   & 
${-5276}_{-2202}^{+1560}$   & ${-5079}_{-1988}^{+1435}$   & 
${-5618}_{-2395}^{+1695}$   & ${-5330}_{-2401}^{+1649}$   & 
${-5011}_{-2080}^{+1487}$   & ${-5686}_{-2596}^{+1820}$   \\
\hline \hline
\end{tabular}
\end{table*}

\begin{table*}[tb]
\caption{
Key quantities of compact stars from the posterior distributions for the 
different astrophysical scenarios: radii, tidal deformabities, central 
baryonic densities, energy densities, pressures, sound speeds, and trace 
anomalies for $1.0,\,1.4,\,2.0\,M_{\odot}$ and the maximum-mass stars. 
We also list the inferred radius differences between $1.4,\,2.0\,M_{\odot}$ 
and $1.0,\,1.4\,M_{\odot}$ stars. The upper and lower values correspond 
to the 95.4\% CI.}
\setlength{\tabcolsep}{2.2pt}
\label{tab:NS_Posterior}
\centering
\begin{tabular}{ccccccccc}
\hline \hline 
 Par.                           & Unit                           & 
 Baseline                       & Scenario A                     & 
 Scenario B                     & Scenario C                     &   
 Scenario D                     & Scenario E                     & 
 Scenario F                     \\
\hline    
 $R_{1.0}$                      & km                             & 
 ${12.53}_{-0.57}^{+0.51}$      & ${12.51}_{-0.44}^{+0.38}$      &
 ${12.44}_{-0.43}^{+0.38}$      & ${12.57}_{-0.41}^{+0.37}$      & 
 ${12.62}_{-0.50}^{+0.45}$      & ${12.53}_{-0.49}^{+0.44}$      & 
 ${12.69}_{-0.47}^{+0.43}$      \\
 $\Lambda_{1.0}$                &                                & 
 ${3287}_{-844}^{+974}$         & ${3233}_{-655}^{+705}$         & 
 ${3114}_{-619}^{+667}$         & ${3340}_{-635}^{+717}$         &  
 ${3413}_{-779}^{+885}$         &${3257}_{-727}^{+824}$          & 
 ${3554}_{-764}^{+858}$         \\
 $\rho_{1.0}$                   & fm$^{-3}$                      & 
 ${0.330}_{-0.050}^{+0.058}$    & ${0.334}_{-0.041}^{+0.045}$    &
 ${0.342}_{-0.041}^{+0.044}$    & ${0.327}_{-0.040}^{+0.042}$    &  
 ${0.325}_{-0.045}^{+0.050}$    & ${0.335}_{-0.045}^{+0.049}$    & 
 ${0.316}_{-0.040}^{+0.047}$    \\
 $P_{1.0}$                      & MeV/fm$^{3}$                   &
 ${28.91}_{-5.61}^{+7.40}$      & ${29.36}_{-4.54}^{+5.68}$      & 
 ${30.27}_{-4.52}^{+5.69}$      & ${28.57}_{-4.34}^{+5.16}$      &  
 ${28.15}_{-4.93}^{+6.25}$      & ${29.30}_{-4.99}^{+6.24}$      & 
 ${27.19}_{-4.43}^{+5.69}$      \\
 $\varepsilon_{1.0}$            & MeV/fm$^{3}$                   & 
 ${328.38}_{-51.61}^{+59.78}$   & ${332.78}_{-43.22}^{+47.33}$   & 
 ${340.87}_{-42.40}^{+45.95}$   & ${325.31}_{-41.17}^{+43.88}$   &  
 ${323.00}_{-46.86}^{+51.72}$   & ${333.20}_{-46.71}^{+50.85}$   & 
 ${313.91}_{-41.88}^{+48.94}$   \\
 $c_{s, 1.0}^2$                 &                                & 
 ${0.269}_{-0.042}^{+0.030}$    & ${0.265}_{-0.039}^{+0.032}$    &
 ${0.263}_{-0.037}^{+0.033}$    & ${0.269}_{-0.041}^{+0.030}$    &  
 ${0.265}_{-0.042}^{+0.032}$    & ${0.262}_{-0.039}^{+0.034}$    & 
 ${0.269}_{-0.044}^{+0.030}$    \\
 $\Delta_{1.0}$                 &                                & 
 ${0.245}_{-0.006}^{+0.005}$    & ${0.245}_{-0.005}^{+0.004}$    & 
 ${0.244}_{-0.005}^{+0.004}$    & ${0.245}_{-0.004}^{+0.004}$    &  
 ${0.246}_{-0.005}^{+0.004}$    & ${0.245}_{-0.005}^{+0.004}$    & 
 ${0.247}_{-0.005}^{+0.004}$   \\
\hline  
 $R_{1.4}$                      & km                             & 
 ${12.61}_{-0.67}^{+0.65}$      & ${12.56}_{-0.52}^{+0.51}$      &
 ${12.47}_{-0.50}^{+0.48}$      & ${12.65}_{-0.49}^{+0.49}$      &  
 ${12.69}_{-0.60}^{+0.59}$      & ${12.57}_{-0.57}^{+0.57}$      &
 ${12.79}_{-0.56}^{+0.55}$      \\
 $\Lambda_{1.4}$                &                                & 
 ${515}_{-167}^{+235}$          & ${500}_{-133}^{+178}$          & 
 ${472}_{-121}^{+163}$          & ${526}_{-134}^{+180}$          &  
 ${535}_{-157}^{+218}$          & ${498}_{-141}^{+198}$          &
 ${571}_{-162}^{+207}$          \\
 $\rho_{1.4}$                   & fm$^{-3}$                      & 
 ${0.405}_{-0.077}^{+0.087}$    & ${0.412}_{-0.067}^{+0.073}$    &
 ${0.425}_{-0.067}^{+0.070}$    & ${0.400}_{-0.063}^{+0.071}$    &  
 ${0.399}_{-0.071}^{+0.078}$    & ${0.415}_{-0.073}^{+0.076}$    & 
 ${0.384}_{-0.062}^{+0.078}$    \\
 $P_{1.4}$                      & MeV/fm$^{3}$                   & 
 ${56.090}_{-14.503}^{+19.278}$ & ${57.537}_{-12.683}^{+15.523}$ & 
 ${60.042}_{-12.859}^{+15.313}$ & ${55.151}_{-11.782}^{+14.406}$ &  
 ${54.680}_{-13.209}^{+16.554}$ & ${57.871}_{-13.733}^{+16.524}$ & 
 ${51.927}_{-11.458}^{+15.549}$ \\
 $\varepsilon_{1.4}$            & MeV/fm$^{3}$                   & 
 ${412.39}_{-83.02}^{+95.53 }$  & ${420.27}_{-73.15}^{+80.62}$   & 
 ${433.76}_{-73.38}^{+76.56}$   & ${406.96}_{-67.94}^{+78.11}$   & 
 ${405.70}_{-76.67}^{+86.29}$   & ${423.26}_{-79.02}^{+82.78}$   &
 ${389.96}_{-66.82}^{+85.26}$   \\
 $c_{s, 1.4}^2$                 &                                & 
 ${0.378}_{-0.050}^{+0.028}$    & 
 ${0.375}_{-0.048}^{+0.028}$    & ${0.374}_{-0.046}^{+0.030}$    & 
 ${0.377}_{-0.049}^{+0.026}$    & ${0.373}_{-0.052}^{+0.029}$    & 
 ${0.372}_{-0.050}^{+0.031}$    & ${0.376}_{-0.054}^{+0.026}$    \\
 $\Delta_{1.4}$                 &                                & 
 ${0.197}_{-0.013}^{+0.010}$    & ${0.197}_{-0.010}^{+0.008}$    & 
 ${0.195}_{-0.010}^{+0.008}$    & ${0.198}_{-0.009}^{+0.008}$    &  
 ${0.199}_{-0.011}^{+0.009}$    & ${0.197}_{-0.011}^{+0.009}$    &
 ${0.200}_{-0.010}^{+0.008}$   \\
\hline  
 $R_{2.0}$                      & km                             & 
 ${12.31}_{-1.24}^{+1.07}$      & ${12.21}_{-1.09}^{+0.92}$      & 
 ${12.04}_{-1.03}^{+0.91}$      & ${12.38}_{-1.05}^{+0.86}$      & 
 ${12.40}_{-1.16}^{+1.00}$      & ${12.17}_{-1.11}^{+1.01}$      & 
 ${12.60}_{-1.14}^{+0.88}$      \\
 $\Lambda_{2.0}$                &                                & 
 ${40.3}_{-25.0}^{+42.6}$       & ${37.6}_{-22.0}^{+33.8}$       & 
 ${33.2}_{-19.1}^{+30.6}$       & ${42.4}_{-23.9}^{+33.9}$       & 
 ${42.42}_{-25.42}^{+40.84}$    & ${36.1}_{-21.4}^{+37.0}$       & 
 ${48.9}_{-28.6}^{+38.6}$       \\
 $\rho_{2.0}$                   & fm$^{-3}$                      & 
 ${0.586}_{-0.162}^{+0.290}$    & ${0.603}_{-0.151}^{+0.272}$    & 
 ${0.636}_{-0.162}^{+0.273}$    & ${0.572}_{-0.134}^{+0.248}$    & 
 ${0.574}_{-0.151}^{+0.273}$    & ${0.616}_{-0.167}^{+0.278}$    & 
 ${0.539}_{-0.126}^{+0.254}$    \\
 $P_{2.0}$                      & MeV/fm$^{3}$                   & 
 ${167.05}_{-70.31}^{+186.55}$  & ${176.17}_{-68.20}^{+176.25}$  & 
 ${193.66}_{-76.50}^{+185.95}$  & ${160.66}_{-58.08}^{+147.96}$  & 
 ${161.06}_{-64.43}^{+168.01}$  & ${182.01}_{-75.67}^{+185.13}$  & 
 ${144.69}_{-51.71}^{+143.77}$  \\
 $\varepsilon_{2.0}$            & MeV/fm$^{3}$                   & 
 ${639.98}_{-195.40}^{+398.34}$ &  
 ${662.57}_{-185.61}^{+376.71}$ & ${704.78}_{-201.48}^{+385.85}$ & 
 ${623.64}_{-162.73}^{+336.38}$ & ${626.21}_{-182.44}^{+371.34}$ &
 ${679.07}_{-206.29}^{+387.85}$ & ${582.67}_{-150.38}^{+339.60}$ \\ 
 $c_{s, 2.0}^2$                 &                                & 
 ${0.568}_{-0.039}^{+0.063}$    & ${0.571}_{-0.035}^{+0.054}$    & 
 ${0.577}_{-0.036}^{+0.057}$    & ${0.565}_{-0.032}^{+0.045}$    & 
 ${0.563}_{-0.038}^{+0.054}$    & ${0.570}_{-0.039}^{+0.059}$    & 
 ${0.557}_{-0.036}^{+0.045}$    \\
 $\Delta_{2.0}$                 &                                & 
 ${0.072}_{-0.080}^{+0.043}$    & ${0.068}_{-0.074}^{+0.040}$    & 
 ${0.059}_{-0.074}^{+0.042}$    & ${0.076}_{-0.064}^{+0.035}$    & 
 ${0.076}_{-0.073}^{+0.040}$    & ${0.065}_{-0.076}^{+0.044}$    & 
 ${0.085}_{-0.065}^{+0.033}$    \\
\hline
 $M_{\rm max}$                  & $M_{\odot}$                    & 
 ${2.26}_{-0.21}^{+0.28}$       & ${2.24}_{-0.19}^{+0.24}$       & 
 ${2.20}_{-0.17}^{+0.23}$       &${2.27}_{-0.21}^{+0.23}$        &   
 ${2.27}_{-0.22}^{+0.27}$       &${2.22}_{-0.19}^{+0.26}$        &
 ${2.32}_{-0.24}^{+0.24}$       \\
 $R_{M_{\rm max}}$              &  km                            & 
 ${11.20}_{-0.76}^{+1.01}$      & ${11.13}_{-0.64}^{+0.84}$      &
 ${11.00}_{-0.58}^{+0.80}$      & ${11.26}_{-0.67}^{+0.82}$      &   
 ${11.26}_{-0.72}^{+0.96}$      & ${11.09}_{-0.64}^{+0.91}$      & 
 ${11.44}_{-0.78}^{+0.88}$      \\
 $\Lambda_{M_{\rm max}}$        &                                & 
 ${5.80}_{-0.85}^{+1.86}$       & ${5.91}_{-0.88}^{+1.89}$       &
 ${6.04}_{-0.93}^{+1.78}$       & ${5.75}_{-0.77}^{+1.92}$       &   
 ${5.83}_{-0.88}^{+2.15}$       & ${6.02}_{-0.98}^{+2.00}$       &
 ${5.64}_{-0.73}^{+2.18}$       \\
 $\rho_{M_{\rm max}}$           &  fm$^{-3}$                     & 
 ${0.955}_{-0.171}^{+0.168}$    & ${0.970}_{-0.152}^{+0.149}$    & 
 ${0.996}_{-0.152}^{+0.136}$    & ${0.943}_{-0.142}^{+0.152}$    &   
 ${0.945}_{-0.162}^{+0.162}$    & ${0.979}_{-0.166}^{+0.145}$    & 
 ${0.912}_{-0.142}^{+0.169}$    \\
 $P_{M_{\rm max}}$              &  MeV/fm$^{3}$                  & 
 ${571.72}_{-76.94}^{+78.72}$   & 
 ${577.90}_{-64.69}^{+64.32}$   &${588.95}_{-62.70}^{+62.09}$    & 
 ${567.45}_{-62.18}^{+62.15}$   & ${564.65}_{-70.65}^{+71.03}$   & 
 ${579.05}_{-69.17}^{+67.76}$   & ${551.63}_{-63.29}^{+69.28}$   \\
 $\varepsilon_{M_{\rm max}}$    &  MeV/fm$^{3}$                  &
 ${1245.03}_{-225.10}^{+220.16}$& ${1265.19}_{-200.68}^{+194.04}$& 
 ${1298.90}_{-199.98}^{+177.49}$& ${1230.27}_{-187.55}^{+197.18}$& 
 ${1232.02}_{-213.08}^{+210.66}$& ${1277.20}_{-218.00}^{+190.11}$& 
 ${1189.94}_{-188.08}^{+219.07}$\\
 $c_{s, M_{\rm max}}^2$         &                                & 
 ${0.734}_{-0.053}^{+0.032}$    & ${0.731}_{-0.053}^{+0.031}$    & 
 ${0.726}_{-0.049}^{+0.032}$    &${0.736}_{-0.055}^{+0.028}$     &  
 ${0.734}_{-0.061}^{+0.032}$    & ${0.727}_{-0.055}^{+0.035}$    & 
 ${0.741}_{-0.063}^{+0.028}$    \\
 $\Delta_{M_{\rm max}}$         &                                & 
 ${-0.130}_{-0.026}^{+0.042}$   & ${-0.127}_{-0.026}^{+0.041}$   & 
 ${-0.123}_{-0.027}^{+0.038}$   & ${-0.131}_{-0.023}^{+0.043}$   &   
 ${-0.129}_{-0.026}^{+0.047}$   & ${-0.124}_{-0.028}^{+0.042}$   & 
 ${-0.134}_{-0.023}^{+0.049}$   \\   
\hline    
 $R_{1.4}$-$R_{2.0}$            & km                             & 
 ${0.28}_{-0.45}^{+0.76}$       & ${0.33}_{-0.44}^{+0.76}$       & 
 ${0.41}_{-0.46}^{+0.74}$       &${0.25}_{-0.39}^{+0.72}$        &    
 ${0.27}_{-0.44}^{+0.77}$       &${0.37}_{-0.48}^{+0.76}$        & 
 ${0.18}_{-0.37}^{+0.75}$       \\
 $R_{1.4}$-$R_{1.0}$            & km                             & 
 ${0.08}_{-0.22}^{+0.22}$       & ${0.06}_{-0.21}^{+0.22}$       & 
 ${0.03}_{-0.20}^{+0.21}$       & ${0.09}_{-0.22}^{+0.20}$       &   
 ${0.08}_{-0.23}^{+0.22}$       & ${0.04}_{-0.21}^{+0.22}$       & 
 ${0.12}_{-0.24}^{+0.20}$       \\
\hline \hline
\end{tabular}
\end{table*}

\begin{table*}[tb]
\caption{
Parameter values for meson-nucleon couplings and their density dependence 
of CDFs from the posterior distributions for the different astrophysical 
scenarios. The upper and lower values correspond to the 68.3\% CI. 
The particle masses adopted are same as those 
in~\citep{Lalazissis:2005,Lijj:2023b}.
}
\setlength{\tabcolsep}{2.2pt}
\label{tab:CDF_Posterior}
\centering
\begin{tabular}{ccccccccc}
\hline \hline 
 Par.                               & Baseline                          &    
 Scenario A                         & Scenario B                        & 
 Scenario C                         & Scenario D                        & 
 Scenario E                         & Scenario F                        \\
\hline 
 $g_{\sigma}$                       & ${9.96267}_{-0.53389}^{+0.61333}$  & 
 ${9.92663}_{-0.51179}^{+0.62975}$  & ${9.88753}_{-0.53912}^{+0.65166}$  & 
 ${9.96500}_{-0.49946}^{+0.60741}$  & ${10.00233}_{-0.52274}^{+0.61930}$ & 
 ${9.94190}_{-0.54769}^{+0.63445}$  & ${10.04090}_{-0.48718}^{+0.59946}$ \\
 $g_{\omega}$                       & ${12.18257}_{-0.79553}^{+0.88983}$ & 
 ${12.12960}_{-0.76465}^{+0.91478}$ & ${12.06945}_{-0.80571}^{+0.95002}$ & 
 ${12.18376}_{-0.74216}^{+0.88206}$ & ${12.23788}_{-0.77561}^{+0.89952}$ & 
 ${12.14935}_{-0.81874}^{+0.92510}$ & ${12.29409}_{-0.71885}^{+0.87200}$ \\
 $g_{\rho}$                         & ${3.60052}_{-0.20111}^{+0.17951}$  & 
 ${3.61178}_{-0.20432}^{+0.17926}$  & ${3.61896}_{-0.20712}^{+0.18320}$  & 
 ${3.60001}_{-0.20211}^{+0.18207}$  & ${3.61310}_{-0.20781}^{+0.18343}$  & 
 ${3.62498}_{-0.20470}^{+0.17850}$  & ${3.60343}_{-0.20629}^{+0.18182}$ \\
\hline 
 $a_{\sigma}$                       & ${1.37785}_{-0.07575}^{+0.10933}$ & 
 ${1.36850}_{-0.06810}^{+0.09765}$  & ${1.36845}_{-0.06421}^{+0.08465}$ & 
 ${1.37376}_{-0.07321}^{+0.11451}$  & ${1.36056}_{-0.07101}^{+0.11155}$ & 
 ${1.35816}_{-0.06621}^{+0.09134}$  & ${1.36740}_{-0.08033}^{+0.13345}$ \\
 $b_{\sigma}$                       & ${0.57954}_{-0.29598}^{+0.99226}$ & 
 ${0.51077}_{-0.25014}^{+0.80903}$  & ${0.45761}_{-0.20903}^{+0.64366}$ & 
 ${0.59951}_{-0.31428}^{+1.01785}$  & ${0.54283}_{-0.28715}^{+0.97983}$ & 
 ${0.45464}_{-0.21850}^{+0.71434}$  & ${0.65861}_{-0.37666}^{+1.30231}$ \\
 $c_{\sigma}$                       & ${0.94443}_{-0.46813}^{+1.59923}$ & 
 ${0.83475}_{-0.39577}^{+1.28249}$  & ${0.75779}_{-0.33366}^{+1.00423}$ & 
 ${0.96720}_{-0.49289}^{+1.65338}$  & ${0.87042}_{-0.44473}^{+1.57783}$ & 
 ${0.74575}_{-0.34750}^{+1.10643}$  & ${1.04371}_{-0.58378}^{+2.13064}$ \\
 $d_{\sigma}$                       & ${0.59409}_{-0.23209}^{+0.24248}$ & 
 ${0.63192}_{-0.23513}^{+0.23948}$  & ${0.66323}_{-0.22829}^{+0.22328}$ & 
 ${0.58706}_{-0.23041}^{+0.25126}$  & ${0.61883}_{-0.24985}^{+0.26606}$ & 
 ${0.66856}_{-0.24434}^{+0.24630}$  & ${0.56513}_{-0.24108}^{+0.28619}$ \\
 $a_{\omega}$                       & ${1.39328}_{-0.09414}^{+0.12045}$ & 
 ${1.38285}_{-0.08885}^{+0.10860}$  & ${1.38177}_{-0.08375}^{+0.09660}$ & 
 ${1.38960}_{-0.09184}^{+0.12545}$  & ${1.37577}_{-0.09186}^{+0.12222}$ & 
 ${1.37168}_{-0.08742}^{+0.10384}$  & ${1.38367}_{-0.09998}^{+0.14757}$ \\
 $b_{\omega}$                       & ${0.58114}_{-0.32471}^{+1.50296}$ & 
 ${0.50515}_{-0.26753}^{+1.22834}$  & ${0.45610}_{-0.23265}^{+1.01049}$ & 
 ${0.60143}_{-0.34050}^{+1.54412}$  & ${0.54736}_{-0.31193}^{+1.49076}$ & 
 ${0.45308}_{-0.23701}^{+1.08372}$  & ${0.66552}_{-0.40193}^{+1.92297}$ \\
 $c_{\omega}$                       & ${0.94128}_{-0.48712}^{+2.35744}$ & 
 ${0.82851}_{-0.40726}^{+1.88498}$  & ${0.75653}_{-0.35083}^{+1.51395}$ & 
 ${0.96860}_{-0.51318}^{+2.45403}$  & ${0.87941}_{-0.46819}^{+2.29319}$ & 
 ${0.74542}_{-0.36095}^{+1.61677}$  & ${1.04460}_{-0.59588}^{+3.12799}$ \\
 $d_{\omega}$                       & ${0.59509}_{-0.27720}^{+0.26163}$ & 
 ${0.63430}_{-0.28381}^{+0.25525}$  & ${0.66378}_{-0.28062}^{+0.24265}$ & 
 ${0.58663}_{-0.27456}^{+0.26889}$  & ${0.61566}_{-0.29152}^{+0.28467}$ & 
 ${0.66871}_{-0.29306}^{+0.26241}$  & ${0.56489}_{-0.28225}^{+0.29700}$ \\
 $a_{\rho}$                         & ${0.57179}_{-0.09381}^{+0.10605}$ & 
 ${0.56209}_{-0.08985}^{+0.10469}$  & ${0.55821}_{-0.08862}^{+0.10244}$ & 
 ${0.57125}_{-0.09389}^{+0.10743}$  & ${0.55872}_{-0.09529}^{+0.10873}$ & 
 ${0.55024}_{-0.08917}^{+0.10437}$  & ${0.56482}_{-0.09668}^{+0.11375}$ \\
\hline\hline
\end{tabular}
\end{table*}

\section{Conclusions}
\label{sec:Conclusions}
Our analysis demonstrates that Bayesian inference applied to CDF models
with density-dependent couplings successfully reconciles recent astrophysical
observations with theoretical predictions for dense nuclear matter. The
$M$-$R$ and $M$-$\Lambda $ distributions derived under various observational
scenarios reveal that maximum mass measurements, tidal deformability constraints
from GW events, and low-density properties constrained by $\chi $EFT play
crucial roles in determining the EOS for CSs. Notably, the softer EOS models,
favored under scenario B, predict lower radii and TDs for canonical-mass
stars compared to the stiffer EOS models favored under scenario F, while
remaining consistent with NICER and multimessenger data. This balance of
constraints illustrates the robustness of CDF models in addressing the
diverse observational data for CSs.

Importantly, the inclusion of updated NICER constraints for pulsars such
as PSR J0437 and J1231 refines the posterior distributions, narrowing the
credible regions of EOS parameters, particularly for sub-canonical and
canonical-mass stars. The results highlight that the stiffness of the EOS
at high densities, driven by isoscalar skewness $Q_{\text {sat }}$, governs
the maximum masses and TDs of massive CSs, with implications for interpreting
GW190814's secondary component as a static CS. Overall, this work provides
tighter constraints on the EOS, advancing our understanding of dense nuclear
matter and its behavior under extreme conditions. The approach underscores
the importance of combining multimessenger observations and advanced theoretical
frameworks to enhance the fidelity of nuclear astrophysics models.

\nolinenumbers
\section*{Acknowledgments}
J.L. and Y.T. acknowledge the support of the National Natural Science 
Foundation of China under Grant No. 12105232 and No. 12475150.  
A. S. is funded by Deutsche Forschungsgemeinschaft Grant No. SE 
1836/6-1 and the Polish NCN Grant No. 2023/51/B/ST9/02798.

\section*{Declaration of competing interest}
The authors declare that they have no known competing financial interests
or personal relationships that could have appeared to influence the work
reported in this paper.

\section*{Data availability}
Data will be made available on request.

\begin{appendices}
\section*{Appendix: \\ 
Key quantities of compact stars and nuclear matter,
and the parameters for underlying CDFs} 

In this appendix, we present the characteristic parameters of symmetric 
nuclear matter at saturation density and key gross quantities of CSs 
predicted by CDFs under seven different scenarios in 
Tables~\ref{tab:NM_Posterior} and~\ref{tab:NS_Posterior}, respectively. 
The parameter values for underlying CDFs are given in 
Table~\ref{tab:CDF_Posterior}. In Table~\ref{tab:NM_Posterior} we also 
show those values for higher order parameters $Z_{\rm sat}$ for isoscalar 
sector and $K_{\rm sym}$, $Q_{\rm sym}$ and $Z_{\rm sym}$ for isovector 
sector which are not allowed to vary freely and thus are the predictions 
of CDFs, once if the low-order characteristics are determined. 
\end{appendices}


\end{document}